\documentclass[aps,nofootinbib,prd,superscriptaddress,eqsecnum,showkeys,preprintnumbers,altaffilletter]{revtex4}
\usepackage{amsfonts,amsmath,amssymb,amsthm,bbm,hyperref}
\usepackage{graphicx}
\usepackage{color}
\usepackage[T1]{fontenc}
\usepackage[normalem]{ulem}

\makeatletter

\def\D{\mathrm{d}}
\def\I{\mathrm{i}}
\def\E{\mathrm{e}}
\def\del{\partial}
\def\be{\begin{equation}}
\def\ee{\end{equation}}

\newcommand{\MP}{M_\mathrm{P}}

\newcommand{\beq}{\begin{eqnarray}}
\newcommand{\eeq}{\end{eqnarray}}
\newcommand{\nn}{\nonumber}

\newcommand{\HdS}{H_\mathrm{dS}}

\makeatother

\begin{document}

\title{Pre-inflation from the multiverse: \\
Can it solve the quadrupole problem in the cosmic microwave background?}

\author{Jo\~{a}o Morais}
\email{jviegas001@ikasle.ehu.eus}
\affiliation{Department of Theoretical Physics, University of the Basque Country UPV/EHU, P.O.~Box 644, 48080 Bilbao, Spain}

\author{Mariam Bouhmadi-L\'{o}pez}
\email{mariam.bouhmadi@ehu.eus}
\affiliation{Department of Theoretical Physics, University of the Basque Country UPV/EHU, P.O.~Box 644, 48080 Bilbao, Spain}
\affiliation{IKERBASQUE, Basque Foundation for Science, 48011 Bilbao, Spain}

\author{Manuel Kr{\"a}mer}
\email{manuel.kraemer@usz.edu.pl}
\affiliation{Institute of Physics, University of Szczecin, Wielkopolska 15, 70-451 Szczecin, Poland}

\author{Salvador Robles-P\'{e}rez}
\email{salvarp@imaff.cfmac.csic.es}
\affiliation{Instituto de F\'{\i}sica Fundamental, CSIC, Serrano 121, 28006 Madrid, Spain}
\affiliation{Estaci\'{o}n Ecol\'{o}gica de Biocosmolog\'{\i}a, Pedro de Alvarado 14, 06411 Medell\'{\i}n, Spain}

\date{\today}

\begin{abstract}
We analyze a quantized toy model of a universe undergoing eternal inflation using a quantum-field-theoretical formulation of the Wheeler--DeWitt equation. This so-called third quantization method leads to the picture that the eternally inflating universe is converted to a multiverse in which sub-universes are created and exhibit a distinctive phase in their evolution before reaching an asymptotic de Sitter phase. From the perspective of one of these sub-universes, we can thus analyze the pre-inflationary phase that arises naturally. Assuming that our observable universe is represented by one of those sub-universes, we calculate how this pre-inflationary phase influences the power spectrum of the cosmic microwave background (CMB) anisotropies and analyze whether it can explain the observed discrepancy of the power spectrum on large scales, i.e.~the quadrupole issue in the CMB. While the answer to this question is negative in the specific model analyzed here, we point out  a possible resolution of  this issue.
\end{abstract}


\maketitle

%
%

\section{Introduction}

In physics, we have been facing the situation
that we are lacking a fully consistent quantum theory of gravity for more than 80 years. Nevertheless, there have been several proposals for a theory of quantum gravity and we need to decide which one of these is the most promising path to be investigated further. Hence, we need predictions that can be tested by experiment or observation. However, we have to deal with the crucial problem that effects from a theory of quantum gravity are expected to only become dominant at very large energies or very small scales that correspond to the Planck scale, i.\,e.~an energy of $10^{19}$\,GeV corresponding to a length scale of $10^{-35}$\,m. Because of this it is extremely difficult to find sizeable effects of quantum gravity. One thus has to look for scenarios in which high energies or large curvature are involved and in this context either black holes or the early universe come to mind. 

Within the last mentioned scenario, for the very first instants of the evolution of our universe, we have the widely accepted theory of inflation that -- apart from providing a solution to the flatness and horizon problem in our universe -- also gives rise to the structure in the universe we observe today. More importantly, the features of inflation are encoded in the anisotropies of the cosmic microwave background (CMB) radiation that has been measured to a large accuracy by the satellites COBE, WMAP and most recently Planck \cite{Planck}. Inflation is estimated to involve energies of the order of $10^{-5}$ times the Planck energy, far more than particle accelerators that operate at energies of up to $10^{-15}$ times the Planck energy.

The anisotropy spectrum of the cosmic microwave background can be described to a remarkable accuracy with a small set of parameters deduced from inflation. However, in recent data there is still a discrepancy at the largest scales that awaits further explanation: The power at these scales is smaller than expected \cite{Planck}. Given that these largest scales exit the horizon at the earliest during inflation, they are also the ones to be influenced by the highest energies during inflation and thus are most likely to carry information about any quantum-gravitational effect happening at or before the onset of inflation.

Furthermore, any theory of quantum gravity should lead to a new fundamental equation that recovers general relativity in a semiclassical limit. For the Wheeler--DeWitt equation it has been shown that a semiclassical approximation gives rise to both general relativity and quantum field theory in curved spacetime and in a subsequent step leads to a functional Schr\"{o}dinger equation with quantum-gravitational corrections \cite{singh}. For an inflationary model these effects can be estimated to be of order $10^{-10}$ and concrete calculations have confirmed this estimate \cite{semi12,semi13,ktv,bkk}.
The magnitude of these effects in thus larger than in other less energetic scenarios, however, due to the fact that these corrections have this magnitude only for large-scale anisotropies -- where also the inherent statistical uncertainty due to cosmic variance is most prominent -- and drop off quickly for smaller scales, it does not seem realistic that such semiclassical effects can be measured -- at least in the CMB power spectrum.  Apart from that it has been discovered that several theories of quantum gravity like Loop Quantum Gravity give rise to a phase that precedes inflation \cite{AAN,SBB16,AG16}. Such a pre-inflationary phase could thus lead to effects that can be orders of magnitude larger than the effects arising from a semiclassical approximation and thus can overcome cosmic variance.

Another aspect when considering inflation is that most inflationary models lead to so-called eternal inflation, which means that the universe as a whole inflates forever while bubbles of spacetime regions form in which inflation eventually ends. These bubbles can be regarded as universes of their own causally disconnected from the others, such that one can speak of a certain kind of multiverse \cite{Linde1983} (see \cite{Everett1957, Susskind2003, Khoury2001, Steinhard2002,Carr2007, Smolin1997, Tegmark2003, Freivogel2004, Mersini2008, mariam2007, RP2007b, Alonso2012} for other models of the multiverse). It has been shown recently that describing the model of an eternally inflating universe using a quantum-field-theoretical formulation of the Wheeler--DeWitt equation -- also called third quantization -- leads to sub-universes exhibiting a phase whose scale factor evolves like $a^{-6}$ before reaching an approximate de Sitter phase \cite{salva2011,Garay:2013pba,Bouhmadi-Lopez:2017sgq}. Hence, from the perspective of a single sub-universe the de Sitter-like inflationary phase is preceded by a pre-inflationary phase. Also modeling an interaction between universes in the third quantization picture leads to specific kinds of pre-inflationary phases \cite{salva2015}.

The objective of this article is thus to analyze a specific model of an inflating universe in the third quantization picture with regard to the effect of the induced pre-inflationary phase on the CMB temperature anisotropies in the sub-universes and to see whether the observed suppression on large scales can be explained by this specific pre-inflationary phase. This is a natural scenario to look for a potential explanation for the low CMB quadrupole, as this pre-inflationary phase will affect the largest modes, i.e.~those that have recently re-entered the horizon and therefore give the largest contribution to the CMB quadrupole. However, before tackling these issues, let us first review what has been done before with regard to this.

There have been different approaches to explain the discrepancy between the theoretical prediction of inflation and the measured temperature anisotropies of the CMB at the largest scales. We list some of them in the following. Fast roll inflation prior to standard inflation \cite{Contaldi:2003zv,Boyanovsky:2006pm}, bounces and cyclic universes \cite{Piao:2003zm,Piao:2005ag,Liu:2013kea}, a radiation dominated era \cite{Powell:2006yg,Wang:2007ws} and a pre-inflationary matter era supported by primordial micro black holes remnants \cite{Scardigli:2010gm} have been considered in the context of the low quadrupole problem. More recently, slow-roll inflation preceded by a topological defect phase \cite{BouhmadiLopez:2012by} as well as compactification before inflation \cite{Kontou:2017xhp} were also suggested as potential ways to explain the low CMB quadrupole.

The article is structured as follows. In the next section, we will present our chosen universe model and explain how we apply the third quantization formalism to it. In section \ref{s-pert}, we will review the methods to calculate the power spectrum of scalar perturbations. In section \ref{Numerical}, we show our results. Then, in section \ref{conclusions}, we present our conclusions. Finally, in Appendix \ref{App_A}, we present further analytical expressions for the cosmological background evolution of the considered model, while in Appendix \ref{App2}, we present an approximation for the evaluation of the power spectrum after horizon crossing.

%
%

\section{Model}
\label{s-model}

We reconsider a model similar to our previous work \cite{Bouhmadi-Lopez:2017sgq} that led to the appearance of an instanton with the crucial difference that we now consider a \emph{flat} Friedmann--Lema\^itre--Robertson--Walker (FLRW) universe with scale factor $a$ instead of a closed one. We shall see that there will be no instanton in this model. As before, we introduce a minimally coupled scalar field $\varphi$ with mass $m$, which follows the potential $\mathcal{V}(\varphi) = \frac{1}{2}\,m^2\varphi^2$.

Canonically quantizing this model leads to the following Wheeler--DeWitt (WDW) equation for the wave function $\phi(a,\varphi)$ \cite{clausbook}
\be \label{basicwdw}
\Biggl[\frac{\hbar^2 G}{3\pi}\,\frac{\del^2}{\del a^2} - \frac{\hbar^2}{4\pi^2 a^2}\,\frac{\del^2}{\del \varphi^2}+2a^4\pi^2 \,\mathcal{V}(\varphi) \Biggr]\phi(a,\varphi) = 0\,.
\ee
We have chosen a specific factor ordering without an additional term containing the first derivative of $\phi$ with respect to $a$, because a term of this type will not have an influence on the subsequent calculations.

The above WDW equation can be simplified using the following rescaling of the scalar field 
\be
\varphi \rightarrow \sqrt{\frac{4\pi G}{3}} \,\varphi\,,
\ee
which absorbs several constants and makes $\varphi$ dimensionless.
We also define the quantities
\be
\label{sigma_def}
H_\varphi^2 := \frac{8 \pi G}{3}\,\mathcal{V}(\varphi) \quad \text{and} \quad \sigma := \frac{3 \pi}{2 G}\,,
\ee
which allows us to introduce
\be
\omega(a,\varphi) := \sigma\,a^2H_\varphi\,.
\ee
Using these definitions, we can thus write the WDW equation \eqref{basicwdw} in the simple form:
\be
\label{parentwdw}
\hbar^2\,\frac{\del^2\phi}{\del a^2} - \frac{\hbar^2}{a^2} \frac{\del^2 \phi}{\del \varphi^2} + \omega^2(a,\varphi) \phi = 0\,.
\ee
We interpret this equation now in the context of eternal inflation and postulate that the creation of  sub-universes during eternal inflation can be described by promoting this wave function $\phi(a,\varphi)$ to an operator $\hat{\phi}(a,\varphi)$ that can be decomposed as follows \cite{salva2011}
\be\label{modedecomp}
\hat{\phi}(a,\varphi) = \int \frac{\D K}{\sqrt{2\pi}} \left[\E^{\I K \varphi} \phi_K(a) \, \hat{b}_{K} + \E^{-\I K \varphi} \phi_K^*(a) \, \hat{b}^\dag_{K}\right].
\ee
The integral is taken over the variable $K$ that specifies the conjugate momentum $p_\varphi$ of the scalar field $\varphi$ for a specific bubble universe. Consequently, the creation and annihilation operators $\hat{b}_{K}$ and $\hat{b}^\dag_{K}$ create and destroy  sub-universes with a specific value of $K$. Each  sub-universe then is described by an amplitude $\phi_K(a)$ that satisfies an effective WDW equation of the form
\be \label{subwdw}
\hbar^2\,\frac{\del^2\phi_K}{\del a^2} + \omega_K^2 \phi_K = 0\,.
\ee
Here, $\omega_K$ is defined as
\be \label{omk}
\omega_K(a) := \sigma \sqrt{a^4 \HdS^2 + \frac{\hbar^2 K^2}{\sigma^2 a^2}}\,.
\ee
We have also introduced a constant $\HdS$ that corresponds to $H_\varphi$ set to the specific value of $\varphi$ (cf. Eq.~\eqref{sigma_def}) that the respective sub-universe takes.

The interpretation here is that this WDW equation \eqref{subwdw} describes the individual sub-universes, in which the scalar field takes a specific value leading to the inflationary scale $\HdS$ and the $\varphi$-derivative term that is present in the parent WDW equation \eqref{parentwdw} leaves its trace as the term $\hbar^2 K^2/a^2$. Note that the factor $\hbar^2$ indicates that this term is clearly of quantum origin.

We thus can now determine what these sub-universes look like, following e.g.~\cite{Garay:2013pba}. An approximate solution to Eq.~\eqref{subwdw} can be described by the Wentzel--Kramers--Brillouin (WKB) ansatz
\be\label{WKB01}
\phi_{\pm,K}(a) \propto \frac{\E^{\pm \frac{\I}{\hbar}\, S_K(a)} }{\sqrt{2 \omega_K(a)}}\,,
\ee
where the function $S_K(a)$ is given by the integral
\be\label{wkbsols}
S_K(a) = \int^a \D \tilde{a} \, \omega_K(\tilde{a})\,,
\ee
The specific linear combination of the solutions (\ref{wkbsols}) will depend on the initial conditions imposed on the wave function of the sub-universe. As it will be clear in the following, we will only assume outgoing modes consistent with an expanding universe.

We now take the WKB solutions \eqref{WKB01} and look at the eigenvalue of the momentum $\hat{p}_a$ at first order given by
\be\label{QMO}
\hat{p}_a \phi_{\pm,K} := -\,\I \,\frac{\partial \phi_{\pm,K}}{\partial a} \approx \pm \,\frac{\partial S_K}{\partial a} \,\phi_{\pm,K} = \pm\,\omega_K \phi_{\pm,K}\,.
\ee
In the semiclassical limit this momentum must be peaked around its classical analogue $p_a = -\,a \frac{\D a}{\D t}$, where $t$ is the cosmic time. Hence, we can deduce that
\be
p_a = -\,a \frac{\D a}{\D t} \approx \mp \,\omega_K(a)\,.
\ee
We only consider the branch with the minus sign in Eq.~(\ref{WKB01}) as this represents an expanding universe. Thus we can finally write down the effective Friedmann equation that describes the evolution of the expanding sub-universes:
\be \label{modFri}
H^2 \equiv \left(\frac{\dot a}{a}\right)^2 = \frac{\omega^2_K(a)}{\sigma^2a^4} = \HdS^2 + \frac{\hbar^2 K^2}{\sigma^2 a^6}\,.
\ee
We can already see at this point that at late times, this universe is asymptotically de Sitter, while for early times a stiff-matter like term $\propto a^{-6}$ appears, which is, however, of pure quantum origin as indicated by the factor $\hbar^2$. Compared to the model analyzed in \cite{Bouhmadi-Lopez:2017sgq} the curvature term is missing (by construction).

Let us now determine the evolution of the scale factor in this sub-universe. 
In order to obtain the explicit solution in terms of the cosmic time, we employ the variable redefinition $a\rightarrow y:=\HdS^3 a^3/ \tilde{K}$ so that the effective Friedmann equation \eqref{modFri} can be written as
\begin{align}
	\label{eq214}
	\D t = 
	\frac{1}{3\HdS}\frac{\D y}{\sqrt{1 + y^2}}
	\,.
\end{align}
Here, we introduce the normalized $K-$number%
\footnote{The normalization employed here is equivalent to $\tilde{K}=2/(3\sqrt{3}) K/K_\mathrm{max}$ where the parameter $K_\textrm{max}$ is defined in \cite{Bouhmadi-Lopez:2017sgq} (the same parameter is defined as $k_m$ in \cite{Garay:2013pba}) as the maximum value of $K$ for which a quantum tunnelling effect can occur in a universe with closed spatial geometry. In the present case, with a flat spatial section, $K_\textrm{max}$ has no particular physical meaning aside from fixing the energy scale of pre-inflation.} 
defined as $\tilde{K}:= (2\hbar^ 2 \HdS^2)/(3\pi\MP^2) K$. Integrating the previous equation from $a=0$ to $a$ we obtain
\begin{align}
	\label{cosmic_time_sol}
	t - t_0 = \frac{1}{3\HdS}\,\mathrm{arsinh}\left(\frac{\HdS^3a^3}{\tilde{K}} \right)
	\,,
\end{align}
where $t_0:=t(a=0)$ can be set to zero without loss of generality. Inverting this equation we find the solution for $a(t)$
\begin{align}
	\label{cosmic_time_sol_a}
	a(t) = \frac{\tilde{K}^{\frac{1}{3}}}{\HdS} \sinh^{\frac{1}{3}}\left[3\HdS\left(t - t_0\right)\right]
	\,.
\end{align}
Alternatively, we can derive the explicit solution of the scale factor in terms of the conformal time $\eta$ ($d\eta:=dt/a$). We begin by employing the variable change $a\rightarrow x:=\HdS^2 a^2/\tilde{K}^{2/3}$, leading to
\begin{align}
	\D\eta = \frac{1}{2\tilde{K}^{1/3}}\frac{\D x}{\sqrt{1 + x^3}}
	\,,
\end{align}
followed by the intermediate transformation%
\footnote{The variable $v$ is defined in the interval $]-1,2-\sqrt{3}]$ where $v(a=0)=2-\sqrt{3}$ and $v(a\rightarrow+\infty)=-1$.}
$x\rightarrow v:=2\sqrt{3}/(1+\sqrt{3}+x)-1$, obtaining
\begin{align}
	\D\eta = -\frac{1}{\tilde{K}^{1/3}}\frac{\D v}{\left\{\left(1-v^2\right)\left[\left(2\sqrt{3}+3\right)v^2 + \left(2\sqrt{3}-3\right)\right]\right\}^{1/2}}
	\,.
\end{align}
A final variable change%
\footnote{The variable $\xi$ is defined in the interval $]0,\arccos(\sqrt{3}-2)]$ where $\xi(a=0)=\arccos(\sqrt{3}-2) $ and $\xi(a\rightarrow+\infty)=0$.}
$v\rightarrow \xi:=\arccos(-v)$ then gives
\begin{align}
	\D\eta = -\frac{1}{2\times3^{1/4}\tilde{K}^{1/3}}
	\frac{\D\xi}{\sqrt{1-\kappa^2\sin^2(\xi)}}
	\,.
\end{align}
where $\kappa^2:=(2+\sqrt{3})/4$.
From the definition of the elliptic integral of the first kind \cite{abra,NIST2010} and after the integration of the previous equation from $a$ to $a\rightarrow+\infty$ we obtain the result
\begin{align}
	\eta_\infty - \eta
	=
	\frac{
	F\left(\xi \middle| \kappa^2 \right)
	}{2\times3^{1/4}\tilde{K}^{1/3}}
	\,,
\end{align}
where $\eta_\infty:=\eta(a\rightarrow+\infty)$ can be set to zero without loss of generality. 
This equality can be inverted using the relation between the elliptic integrals and the Jacobi elliptic functions \cite{abra,NIST2010} giving the solution for $a(\eta)$:
\begin{align}
	\label{conformal_time_sol_a}
	a(\eta)
	=
	\frac{\tilde{K}^{\frac{1}{3}}}{\HdS}
	\left(
		\frac{\left(\sqrt{3}-1\right) + \left(\sqrt{3}+1\right)\mathrm{cn}\left[
			2\times3^{1/4}\tilde{K}^{1/3}\left(\eta_\infty-\eta\right)
			\middle|
			\kappa^2
		\right]}
		{1-\mathrm{cn}\left[
			2\times3^{1/4}\tilde{K}^{1/3}\left(\eta_\infty-\eta\right)
			\middle|
		\kappa^2
		\right]}
	\right)^{\frac{1}{2}}
	\,.
\end{align}

\begin{figure}[t]
\centering
\includegraphics[width=.475\textwidth]{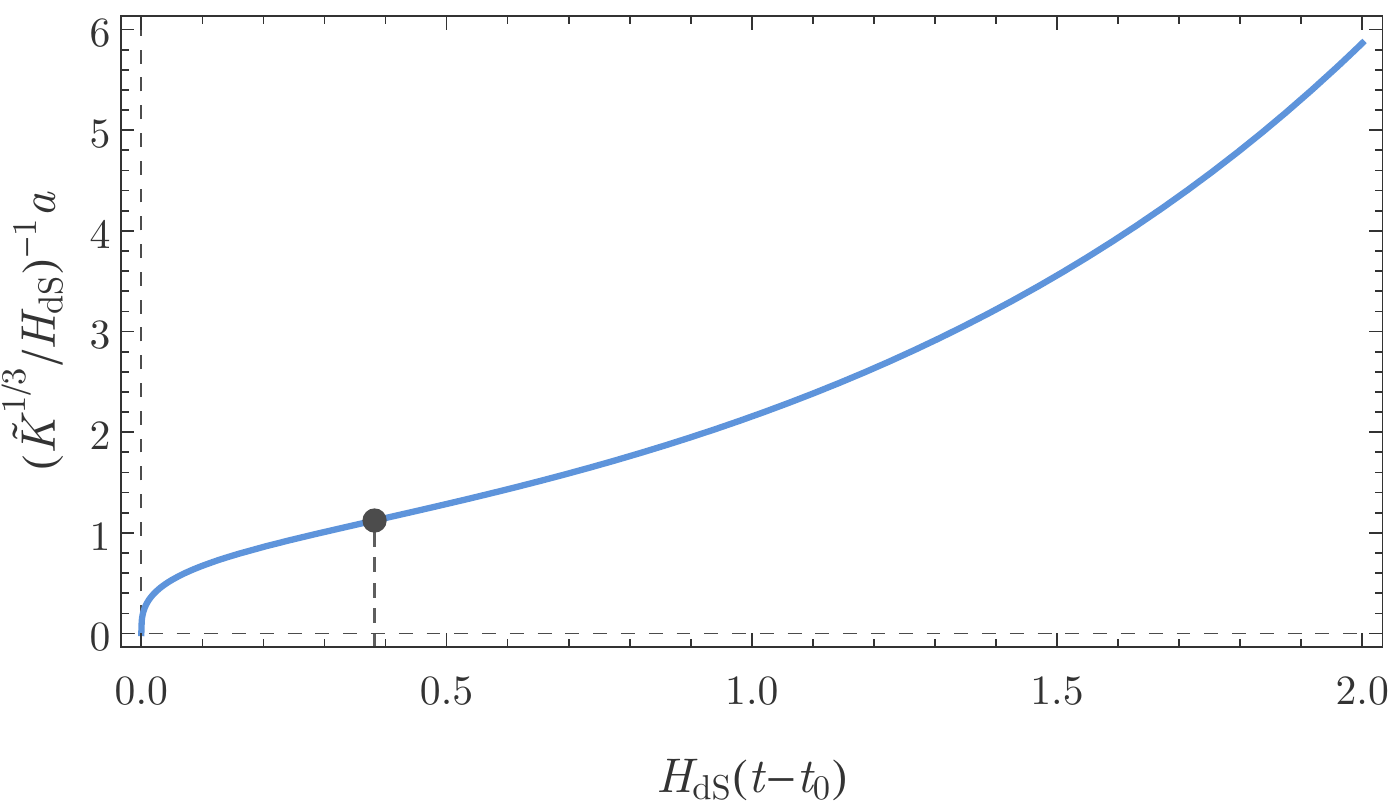}
\hfill
\includegraphics[width=.475\textwidth]{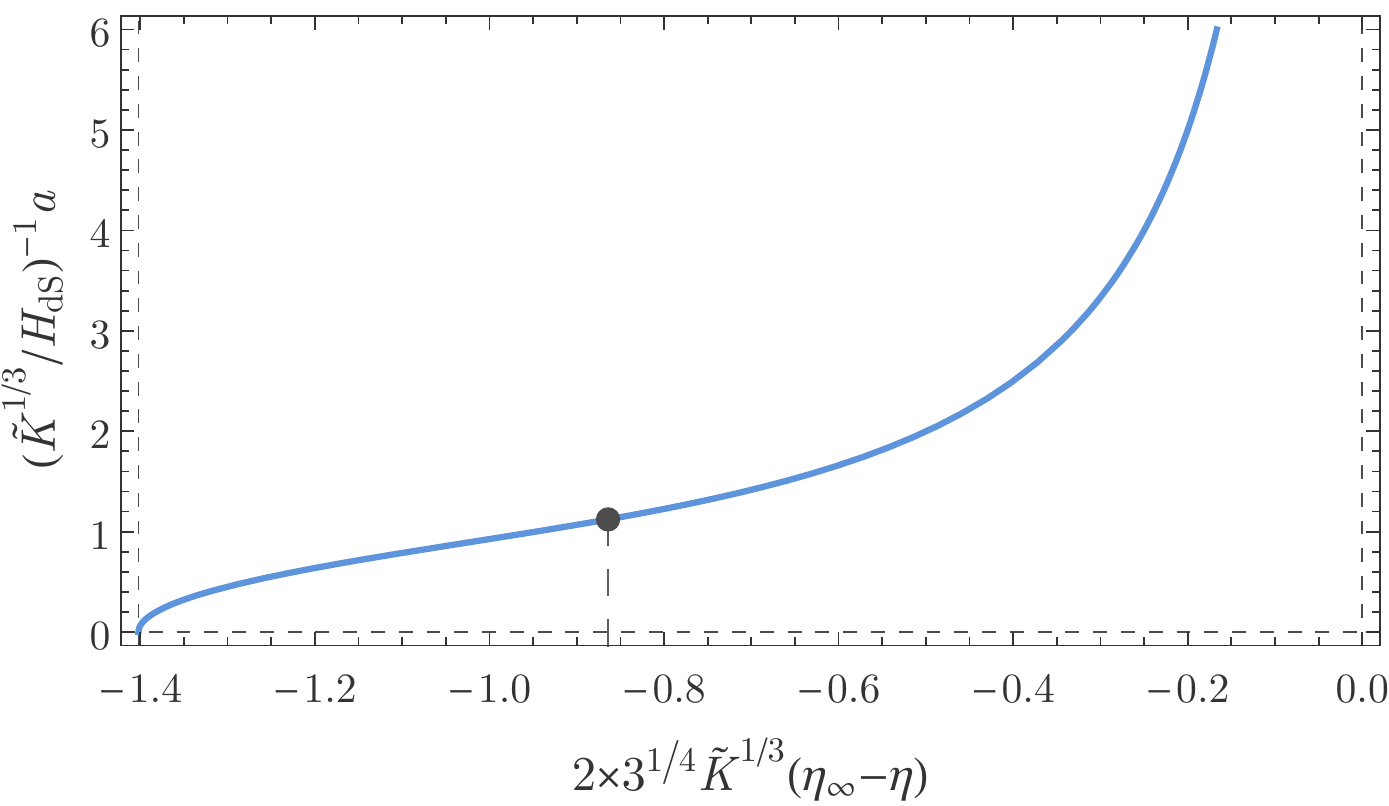}
\caption{\label{fig_initial conditions} The evolution of the scale factor as a function of the cosmic time (cf. Eq.~\eqref{cosmic_time_sol_a}) and of the conformal time (cf. Eq.~\eqref{conformal_time_sol_a}). The moment of transition from decelerated expansion to inflation is indicated by a dark dot.}
\end{figure}

In order to be able to calculate how scalar perturbations evolve in this universe and to eventually determine their power spectrum, we introduce these perturbations as being perturbations of the constant scalar field inherited from the parent universe that leads to the cosmological constant $\HdS$. We are aware that this is just a first step towards a full resolution of the problem.

%
%

\section{Scalar Perturbations}
\label{s-pert}

In a spatially flat FLRW  model, we can write the scalar part of the linearly perturbed line element in terms of the four scalars $A$, $B$, $\psi$ and $E$ as \cite{Bassett:2005xm}
\begin{align}
	\D s^2 = 
	-a^2\left(1+2A\right) \D\eta^2 
	+ 2a^2\partial_iB \,\D x^i \D\eta
	+a^2\left[\left(1-2\psi\right)\delta_{ij} + 2\partial_i\partial_j E\right]\D x^i \D x^j
	\,.
\end{align}
If we consider the presence of a single minimally coupled scalar field $\tilde{\varphi}$, then the dynamics of the scalar sector can be appropriately described in terms of the gauge-invariant Mukhanov--Sasaki variable $v = a[\delta\tilde{\varphi} + (\tilde{\varphi}'/\mathcal{H})\psi]$ \cite{Mukhanov:1990me}, which can be related to the comoving curvature perturbation $\mathcal{R}$ through $v=z\mathcal{R}$ \cite{Lyth:1984gv,Mukhanov:1990me,Bassett:2005xm,Mukhanov:2005sc}. Here, we have used the definitions $(\cdot)':=\D\cdot/\D\eta$, $\mathcal{H}:=a'/a$ and $z= a (\tilde{\varphi}'/\mathcal{H})$. The second-order action for the variable $v$ is \cite{Mukhanov:1990me,Mukhanov:2005sc} 
\begin{align}
	\label{2order_action}
	\delta_2S= \frac{1}{2}\int \D\eta\,\D^3\vec{x} \left[(v')^2 + v\,{\partial^i\partial_ i}v + \frac{z''}{z}v^2\right]
	\,,
\end{align}
which, once minimized, leads to the evolution equation \cite{Mukhanov:1990me,Mukhanov:2005sc}:
\begin{align}
	\label{quantum_evolution}
	{v}'' - \vec{\nabla}{v} - \frac{z''}{z}{v} = 0
	\,.
\end{align}
The potential $z''/z$ that appears as an effective mass term  in the second term of  Eq.~\eqref{2order_action} can be written as \cite{Bassett:2005xm}
\begin{align}
	\frac{z''}{z}
	=
	\left(aH\right)^2\left[
		2 
		+ 2\epsilon 
		- 3\delta 
		+3\epsilon^2
		-5\epsilon\delta
		+\delta^2
		+\xi^2
	\right]
	\,,
\end{align}
where the slow-roll parameters $\epsilon$, $\delta$ and $\xi$ are defined as \cite{Bassett:2005xm,bkk}
\begin{align}
	\epsilon
	:=
	-\frac{\dot{H}}{H^2} 
	= 
	4\pi\frac{\dot{\varphi}^2}{H^2}
	\,,
	\qquad
	\delta
	:=
	\epsilon
	- \frac{\dot{\epsilon}}{2H\epsilon}
	=
	-\frac{\ddot{\varphi}}{H\dot{\varphi}}
	\,,
	\qquad
	\textrm{and}
	\qquad
	\xi 
	:=
	2\epsilon\left(\epsilon+\delta\right) - \frac{\dot{\epsilon}+\dot{\delta}}{H}
	\,.
\end{align}

We can now follow the canonical quantization scheme and elevate the perturbation $v$ to an operator $\hat{v}$ with conjugate momentum $\hat\pi=\hat v'$. These obey the standard commutation relations \cite{Lyth:1984gv,Mukhanov:1990me,Mukhanov:2005sc}
\begin{align}
	\label{quantum_commutations}
	\left[\hat v(\eta,\vec{x}),\hat v(\eta,\vec{y})\right] = 0
	\,,
	\qquad
	\left[\hat \pi(\eta,\vec{x}),\hat \pi(\eta,\vec{y})\right] = 0
	\,,
	\qquad
	\left[\hat v(\eta,\vec{x}),\hat \pi(\eta,\vec{y})\right] = \I\hbar\,\delta\left(\vec x- \vec y\right)
	\,.
\end{align}
while $\hat{v}$ satisfies the same equation \eqref{quantum_evolution} as its classical counterpart. At the linear level in the perturbations, it is advantageous to employ a Fourier decomposition of the operator $\hat{v}$ \cite{Baumann:2009ds}
\begin{align}
	\label{v_Fourier}
	\hat{v} (\eta,\vec{x}) = \int \frac{\D^3\vec{k}}{(2\pi)^{3/2}}\left[
		v_{\vec{k}}(\eta )\,\E^{i \vec{k}\cdot\vec{x}} \hat{a}_{\vec{k}}^-
		+ v_{\vec{k}}^*(\eta )\,\E^{-i \vec{k}\cdot\vec{x}} \hat{a}_{\vec{k}}^+
	\right]
	\,,
\end{align}
where $v_{\vec{k}}(\eta )$ are the mode functions for the scalar perturbations and $\hat a_{\vec k}^-$ and $\hat a_{\vec k}^+$ are annihilation and creation operators that satisfy $[\hat a_{\vec k}^-,\, \hat a_{\vec {k'}}^-]=0$, $[\hat a_{\vec k}^+,\, \hat a_{\vec {k'}}^+]=0$ and $[\hat a_{\vec k}^-,\, \hat a_{\vec {k'}}^+]=\delta({\vec {k}}-{\vec {k'}})$ \cite{Lyth:1984gv,Mukhanov:1990me,Mukhanov:2005sc}.
Inserting the decomposition \eqref{v_Fourier} in Eqs.~\eqref{quantum_evolution} and \eqref{quantum_commutations} we find that each mode function $v_{\vec{k}}$ verifies the normalization relation \cite{Baumann:2009ds}
\begin{align}
	\label{normalization}
	v_{\vec{k}}{v_{\vec{k}}^*}'
	- v_{\vec{k}}^* {v_{\vec{k}}}'
	= \I\hbar
	\,,
\end{align}
while satisfying the equation \cite{Mukhanov:1990me,Bassett:2005xm,Mukhanov:2005sc}
\begin{align}
	\label{mode_evolution}
	v_{\vec{k}}'' + \left(k^2 - \frac{z''}{z}\right)v_{\vec{k}} = 0
	\,,
\end{align}
where $k^2= \vec{k}\cdot \vec{k}$. 
In the small wave-number limit, $k^2\gg z''/z$, the ground state of the Hamiltonian of the system is characterized by the Bunch--Davies vacuum solution \cite{Mukhanov:1990me,Bassett:2005xm,Mukhanov:2005sc}
\begin{align}
	\label{ground_state}
	v_{\vec k} = \sqrt{\frac{\hbar}{2k}}\,\E^{-\I k\eta}
	\,,
\end{align}
which can be checked to satisfy the normalization relation \eqref{normalization}. Once the opposite regime $k^2<z''/z$ is achieved, oscillatory solutions like the Bunch--Davies vacuum are no longer valid and an enhancement or suppression of the amplitude of the mode can occur. 
In the particular case of a constant equation of state (EoS) parameter $w:=P/\rho$, the evolution equation \eqref{mode_evolution} for the modes $v_k$ can be solved analytically in terms of the Hankel functions of the first and second kind \cite{abra,NIST2010} and order $\lambda$ \cite{Bassett:2005xm,Contaldi:2003zv}:
\begin{align}
	\label{solutions_anal}
	v_k = 
	\frac{\sqrt{\pi\hbar|\eta|}}{2}\left[
		c_{1k} H_{\lambda}^{(1)}\left(k|\eta|\right)
		+ c_{2k} H_{\lambda}^{(2)}\left(k|\eta|\right)
	\right]
	\,,
	\qquad
	\lambda:=\lambda(w)= \sqrt{2\dfrac{1-3w}{\left(1+3w\right)^2}
			+ \frac{1}{4}}
	\,,
\end{align}
where%
\footnote{For $w=-1/3$ the potential $z''/z$ is constant, which leads to trivial oscillatory solutions for $v_k$.}
$\eta>0$ for $w>-1/3$ and $\eta<0$ if $w<-1/3$. The linear coefficients $c_{1k}$ and $c_{2k}$ satisfy the constraint
\begin{align}
	\label{cik_normalization}
	|c_{1k}|^2 - |c_{2k}|^2 = \pm 1
	\,,
\end{align}
which is derived from the normalization relation~\eqref{normalization}.
The upper positive sign in \eqref{cik_normalization} corresponds to inflationary phases of $w<-1/3$, while the lower negative sign corresponds to the cases with $w>-1/3$. 
While in a general case no such solutions can be found, by analysing the evolution of $z''/z$ we can understand which range of modes will be affected during a certain period of the expansion of the Universe. 


Usually, quantum zero-point fluctuations of a given quantity are described in terms of their variance $\langle v^2(\eta,\vec{x})\rangle $ or, alternatively, its Fourier transform, the primordial power spectrum $P_v(k)=k^3|v_k|^2/(2\pi^2)$. Using the relation between the Mukhanov--Sasaki variable and the comoving curvature perturbation, we can write the primordial power spectrum for $\mathcal{R}$ as \cite{Bassett:2005xm}
\begin{align}
	\label{power spectrum}
	P_{\mathcal{R}}(k) 
	= \frac{k^3}{2\pi^2}|\mathcal{R}_k|^2
	= \frac{k^3}{2\pi^2}\frac{|v_k|^2}{z^2}
	\,.
\end{align}
Observationally, the primordial power spectrum is usually fitted to a power-law
\begin{align}
	\label{power_spectrum_fit}
	P_{\mathcal{R}}(k) = A_s \left(\frac{k}{k_*}\right)^{n_s-1}
	\,,
\end{align}
where $A_s$ and $n_s$ represent the amplitude and the tilt of the spectrum at the pivot scale $k_*$. The results of the Planck2015 mission in combination with lensing efffects and external data (BAO+JLA+H0) \cite{Planck} fix these parameters at 
$A_s =2.142\pm0.049$ and $n_s=0.9667\pm0.0040$ for a pivot scale $k_*$ corresponding to $0.05$~$\mathrm{Mpc}^{-1}$.

In order to obtain a theoretical prediction for $P_\mathcal{R}$ we need to fix the linear coefficients $c_{ik}$ in \eqref{solutions_anal}. While the normalization condition \eqref{cik_normalization} only gives a relation between their amplitudes, by requiring that \eqref{ground_state} is recovered for $|k\eta|\gg1$, i.e.~in the low wave-length regime of the Mukhanov--Sasaki equation when the modes are inside the Hubble horizon, we can fully fix $c_{1k}$ and $c_{2k}$ minus an arbitrary phase with no impact on the physical results.
In the case of a constant parameter of EoS we can use the asymptotic behavior of the Hankel functions for large values of the argument to obtain $c_{1k}=0$ and $|c_{2k}|=1$ for $w>-1/3$ and $|c_{1k}|=1$ and $c_{2k}=0$ for $w<-1/3$.
With these choices of the coefficients the primordial power spectrum \eqref{power spectrum} becomes
\begin{align}
	\label{Hankel_power}
	P_{\mathcal{R}}(k) 
	= \frac{k^3}{16\pi\epsilon} \frac{\hbar |\eta|}{a^2}\left|H_\lambda^{(1)}(k|\eta|)\right|^2
	\,,
\end{align}
where we have used the property that the two Hankel functions with a real-valued variable are the complex conjugate of each other \cite{abra,NIST2010}.
For long wave-numbers, when the modes are in the ground state, the power spectrum in \eqref{Hankel_power} reduces to \cite{Bassett:2005xm}
\begin{align}
	P_{\mathcal{R}}(k|\eta|\gg1) 
	\simeq
	\frac{1}{\pi\epsilon}\frac{\hbar^2 H^2}{\MP^2} \left(\frac{k}{a H}\right)^2
	\,,
\end{align}
independently of the value of $w$.
In the opposite low wave-number regime ($k|\eta|\ll1$) we find that the shape of the primordial power spectrum depends on the kind of expansion that the universe is undergoing. In the case of $w=1$ the expression in \eqref{Hankel_power} reduces to \cite{abra,NIST2010}
\begin{align}
	\label{spectrum_w1}
	P_{\mathcal{R}}(k|\eta|\ll1) 
	\simeq
	\frac{1}{\pi^2\epsilon}\frac{\hbar^2 H^2}{\MP^2} \left(\frac{k}{a H}\right)^3\log^2\left(\frac{k}{2 a H}\right)
	\,.
\end{align}
Notice that the logarithmic dependence is a consequence of an effective stiff-matter behavior. Likewise, for an arbitrary $w<1$ and in particular for near de Sitter inflation with $w=-1+\alpha/3\gtrsim-1$, we get
\begin{align}
	\label{spectrum_walpha}
	P_{\mathcal{R}}(k|\eta|\ll1) 
	\simeq
	\frac{1}{\pi\epsilon}\left[{\left(1-\alpha/2\right)}\frac{\Gamma\left(\frac{1}{2}\frac{6-\alpha}{2-\alpha}\right)}{\Gamma(3/2)}\right]^2\frac{\hbar^2 H^2}{\MP^2} \left[\frac{k}{(2-\alpha)a H}\right]^{-\frac{2\alpha}{2-\alpha}}
	\,.
\end{align}


\section{Numerical Results}
\label{Numerical}

\subsection{Toy model}
\label{Sec_AsympSols}

As a toy model we relax the asymptotic de Sitter behavior in Eq.~\eqref{modFri} by replacing the constant term $\HdS^ 2$ by a term proportional to $a^{-\alpha}$, where $\alpha\gtrsim0$. 
This is just a first approach and we are aware that a much more realistic approach would be desirable to fit the current bounds for the tensor-to-scalar ratio usually denoted as $r$.
This power-law inflationary behavior introduces a tilt in the primordial power spectrum, thus leading to a value of $n_s$ different from unity. The effective Friedmann equation \eqref{modFri} now becomes
\begin{align}
	\label{bkgd_power_law}
	H^2
	=
	\HdS^2\left[
		\left(\frac{a_*}{a} \right)^{\alpha}
		+ \frac{\tilde{K}^2}{\HdS^6 a^6}
	\right]
	\,,
\end{align}
where $a_*$ is an arbitrary scale, to be fixed later, for which the magnitude of the inflationary term is $\HdS^2$. This equation describes the transition from an initial \textit{quantum} stiff-matter-like epoch with EoS $w=1$, dominated by the $a^{-6}$ term, to a later power-law inflation period with $w=-1 + \alpha/3$. 
At the moment of transition from a decelerated expansion to inflation in  Eq.~\eqref{bkgd_power_law} the value of the scale factor is given by
\begin{align}
	\label{transition}
	a_\mathrm{trans} = a_*\left(\frac{4}{2-\alpha}\frac{\tilde{K}^2}{a_*^6\HdS^6}\right)^{\frac{1}{6-\alpha}}
	\,.
\end{align}
This corresponds to the minimum value of $(aH)^2$ during the transition to inflation, given by
\begin{align}
	\label{kmin_def}
	k^2_\mathrm{min}
	:= \left(aH\right)^2_{a=a_\mathrm{trans}}
	= \frac{6-\alpha}{2-\alpha}\left(\frac{2-\alpha}{4}\right)^{\frac{4}{6-\alpha}}\left(\frac{\tilde{K}^2}{a_*^6\HdS^6}\right)^{\frac{2-\alpha}{6-\alpha}} \left(a_*\HdS \right)^2
	\,.
\end{align}
Notice that in the limit of $\alpha=0$ this equality reduces to $k^2_\mathrm{min}=(27/4)^{1/3}\tilde{K}^{2/3}$, which is consistent with the former Friedmann equation \eqref{modFri}.
While solutions of \eqref{bkgd_power_law} for $t(a)$ and $\eta(a)$ can be found and involve hypergeometric functions \cite{abra,NIST2010}, since the inverse relations $a(t)$ and $a(\eta)$ could not be obtained, we present these solutions only in Appendix~\ref{App_A}.

\begin{figure}[t]
\centering
\includegraphics[width=.475\textwidth]{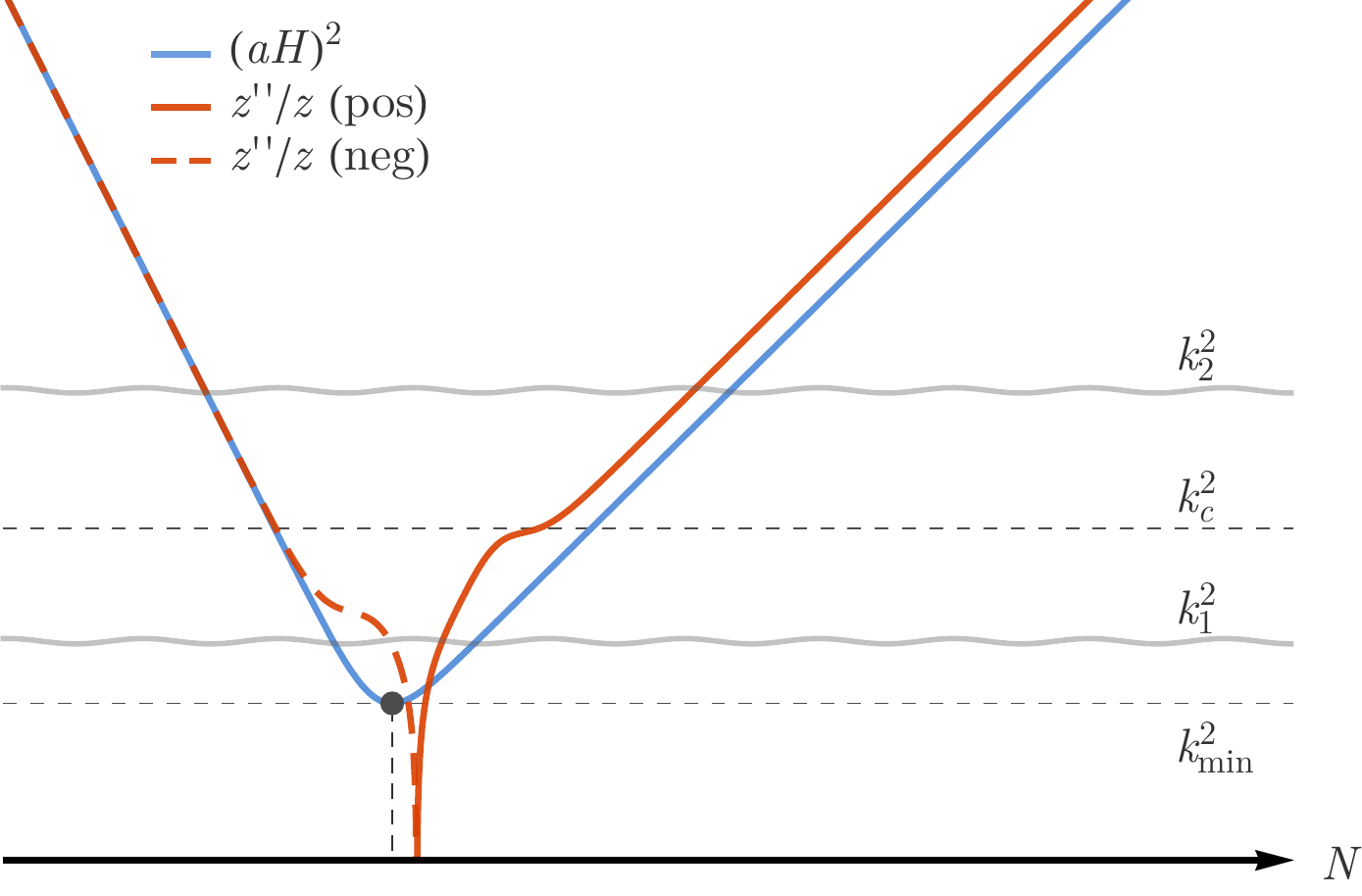}
\hfill
\includegraphics[width=.475\textwidth]{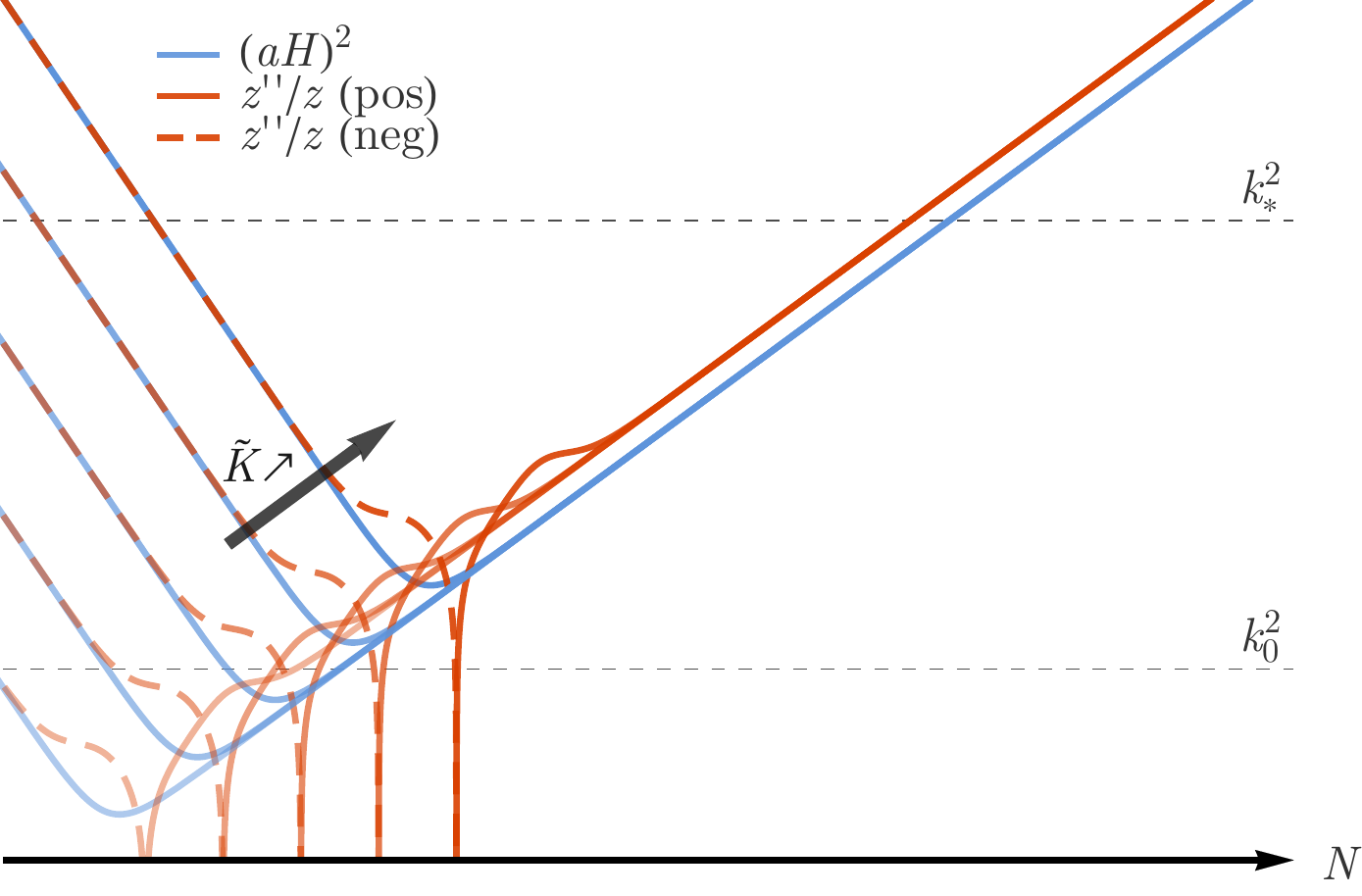}
\caption{\label{fig_potential} (Left Panel) The shape of the potential $z''/z$ (red) and $(aH)^2$ (blue) around the transition from decelerated expansion to power-law inflation. The imprints of the transition on the primordial power spectrum are expected on modes with wave-number $k_1\lesssim k_c$ where $k_c$ is a characteristic scale defined by the features (bumps) of the potential during the transition. For $k_2\gg k_c$ the modes should be ``blind'' to the shape of the potential during inflation and the usual shape of $P_\mathcal{R}$ should be recovered. The moment of transition from decelerated expansion to inflation, when $(aH)^2=k_\mathrm{min}^2$, is indicated by a dark dot. (Right Panel) As $\tilde{K}$ increases the transition from decelerated expansion to power-law inflation affects higher wave-numbers. In order for any imprints of the transition to be visible on the primordial power spectrum the bumps of the potential $z''/z$ should be above $k_0^2$, where $k_0$ corresponds to the Hubble horizon at the present time. However, for the constraints on $P_\mathcal{R}$ around the pivot scale $k_*$ to be satisfied these bumps should be well below the $k_*^2$.}
\end{figure}

On the left-hand side panel of Fig.~\ref{fig_potential} we present the evolution of $z''/z$ (red) and of the wave-number of the Hubble horizon (blue) for the background model in Eq.~\eqref{bkgd_power_law}. 
In the asymptotic regimes, when $z''/z$ tracks the behavior of $(aH)^2$, we find that the mode functions evolve according to the solution \eqref{solutions_anal}, with $\lambda=0$ during the initial ``stiff'' epoch and $\lambda=(6-\alpha)/(4-2\alpha)$ during the subsequent power-law inflation.
In the transition between these two regimes the solution \eqref{solutions_anal} is no longer valid and a numerical integration of the Mukhanov--Sasaki equation is required. We can, however, analyze the features of the potential during such a transition, where we find a reversal of sign accompanied by two ``bumps''. Defining a characteristic wave-number $k_c$ through the maximum value of $z''/z$ in this interval (cf. Fig.~\ref{fig_potential}), we expect to find imprints of the transition in the modes with wave-number $k_1\lesssim k_c$. For wave-numbers $k_2\gg k_c$, we expect the modes to have enough time to achieve the Bunch--Davies vacuum before exiting the Hubble horizon during inflation. As such, these modes should have no memory of the pre-inflationary evolution and are ``blind'' to the shape of the potential during the transition.

Observational data sets strict constraints on the shape of the primordial power spectrum around the pivot scale $k_*$, which is very well fitted by the power-law \eqref{power_spectrum_fit}. This indicates that there should be no pre-inflationary imprints on $P_\mathcal{R}$ for scales near $k_*$ and that we can use the constraints on $A_s$ and $n_s$ to fix the values of the parameters of our toy model -- $(\HdS,a_*,\alpha,\tilde{K})$ -- for numerical computations. 
Evaluating Eq.~\eqref{spectrum_walpha} at the moment of horizon crossing, $k=aH$, we find%
\footnote{While strictly speaking the formula \eqref{spectrum_walpha} is only valid for $k|\eta|\ll1$, i.e. after the mode exits the Hubble horizon, it can be shown that a good approximation to the power spectrum after inflation can be obtained by extending this solution to the moment of horizon crossing, when $k|\eta|\approx1$. For more details please see App.~\ref{App2}.}
\begin{align}
	\label{power_spectrum_approx}
	P_{\mathcal{R}}(k\approx k_*)
	\simeq
	\frac{1}{2\pi\alpha}\left[{\left(2-\alpha\right)}\frac{\Gamma\left(\frac{1}{2}\frac{6-\alpha}{2-\alpha}\right)}{\Gamma(3/2)}\right]^2 \left(\frac{1}{2-\alpha}\right)^{-\frac{2\alpha}{2-\alpha}} \frac{\hbar^2 \HdS^2}{\MP^2}\left(\frac{k}{a_*\HdS}\right)^{-\frac{2\alpha}{2-\alpha}}
	\,,
\end{align}
where we have used the equality $\epsilon=\alpha/2$ valid during power law inflation. By comparing \eqref{power_spectrum_approx} with the observational fit \eqref{power_spectrum_fit} and using the freedom in  choosing  the value of $a_*$ to fix $k_* = a_*\HdS$ we can write the parameters $\HdS$, $a_*$, and $\alpha$ in terms of the cosmological parameters as 
\begin{align}
	\alpha = 2 \frac{1-n_s}{3-n_s}
	\,,
	\quad
	\HdS = \frac{\pi}{2}\frac{\sqrt{\left(1-n_s\right)A_s}}{\Gamma\left(2-n_s/2\right)} \left(\frac{3-n_s}{4}\right)^{1-\frac{n_s}{2}} \frac{\MP}{\hbar}
	\,,
	\quad
	a_* = \frac{2k_*}{\pi} \frac{\Gamma\left(2-n_s/2\right)}{\sqrt{\left(1-n_s\right)A_s}} \left(\frac{3-n_s}{4}\right)^{\frac{n_s}{2}-1} \frac{\hbar }{\MP }
	\,.
\end{align}
Using the constraints of the 2015 data release of the Planck mission in combination with lensing effects and external data (BAO+JLA+H0) \cite{Planck}, we obtain the values
\begin{align}
	\alpha \approx 0.03275
	\,,
	\qquad
	\HdS
	\approx
	1.055\times10^{-5}\,(\MP/\hbar)
	\,,
	\qquad
	 a_*
	\approx
	 2.483\times10^{-54}\,(\hbar/\MP)
	 \,,
\end{align}
while $H_0\approx1.184\times10^{-61}\,(\MP/\hbar)$ and $ a_0=1\,(\hbar/\MP)$.

Having fixed $\HdS$, $a_*$, and $\alpha$, we are left with one free parameter -- $\tilde{K}$ -- that is related to the initial energy density of the momentum $p_\varphi$ and modulates the term $a^{-6}$  in the effective Friedmann equation \eqref{bkgd_power_law}. From Eq.~\eqref{transition} we have $a_\mathrm{trans}\sim \tilde{K}^{2/(6-\alpha)}$, therefore the transition from the decelerated expansion to inflation affects progressively higher wave-numbers as the value of $\tilde{K}$ increases (cf. the right hand side panel of Fig.~\ref{fig_potential}). In order for the imprints on the power spectrum to be visible without violating the constraints around the pivot scale, the bumps of the potential during the transition should be above $k_0^2$, where $k_0$ corresponds to the Hubble horizon at the present time, but far below $k_*^2$. 
Numerical investigation indicates that the value of $k_c$ can be well approximated by
\begin{align}
	k_c \approx \sqrt{50} \tilde{K}^{\frac{1}{3}}
	\,.
\end{align}
Imposing the limits $k_0\lesssim k_c\lesssim10^{-1}k_*$, we can consider that the range of values of $\tilde{K}$ that satisfy the observational constraints coming from the CMB anisotropy spectrum while introducing visible imprints in the primordial power spectrum is
\begin{align}
	\label{define_constraint}
	\left(\frac{k_*}{221\sqrt{50}}\right)^3\lesssim \tilde{K}\lesssim\left(\frac{k_*}{10\sqrt{50}}\right)^3
	\,.
\end{align}
For these values, we can constrain the number of e-foldings of inflation before $a=a_*$ as
\begin{align}
	4.161\lesssim \log\left(\frac{a_*}{a_\mathrm{trans}}\right)\lesssim7.274
	\,.
\end{align}

\subsection{Numerical Computations}

\begin{figure}[t]
\centering
\includegraphics[width=.500\textwidth]{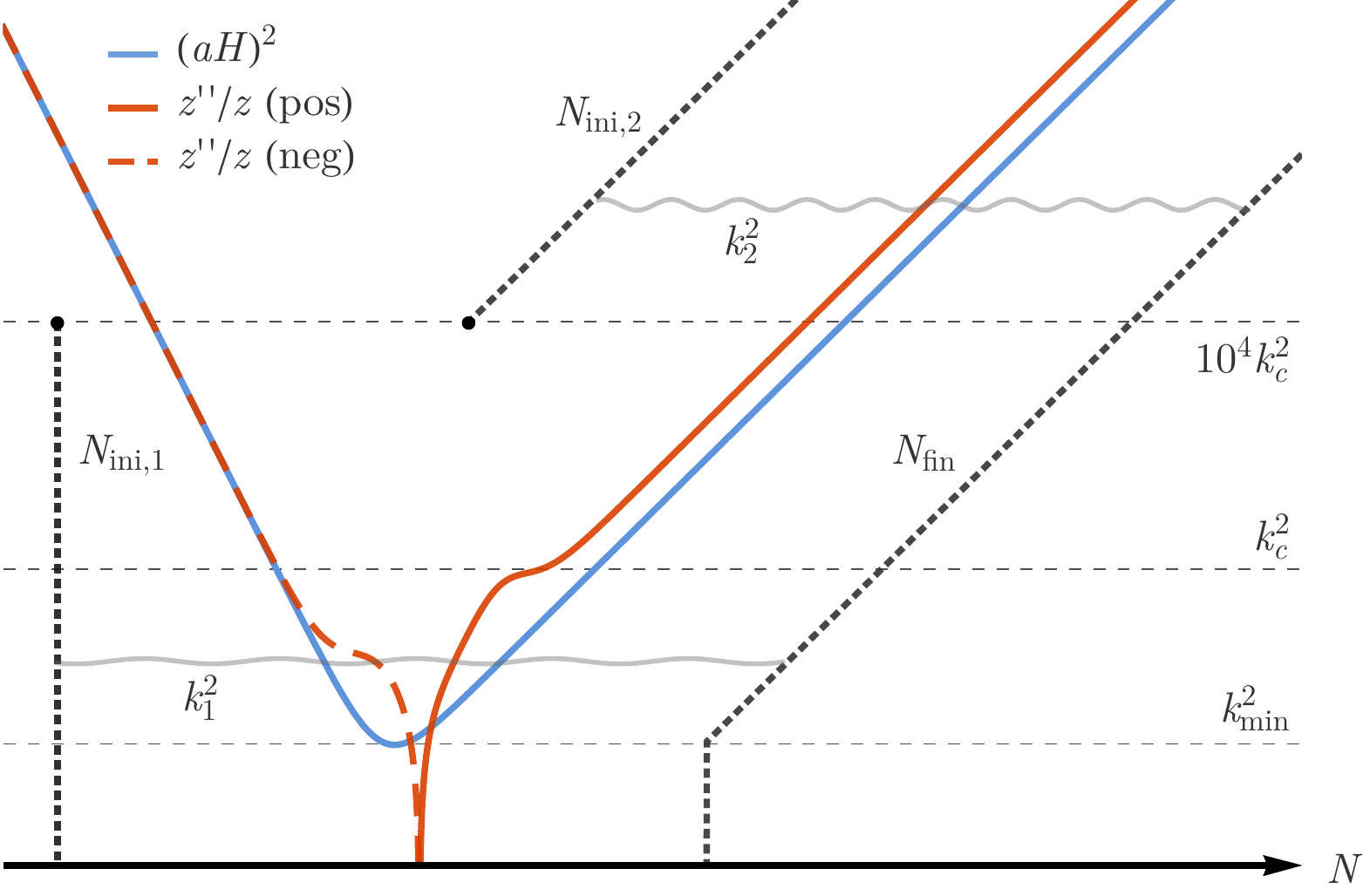}
\caption{\label{fig_initial conditions} The initial conditions for the numerical integration are set for each mode depending on whether the wave-number $k$ is above or below $10^2k_c$, where $k_c$ characterizes the bumps in the potential $z''/z$ around the transition to inflation. For $k_1<10^2k_c$ the initial conditions are set at $N_{\mathrm{ini},1}$ well during the initial ``stiff'' epoch. For $k_2>10^2k_c$ the initial conditions are set some $N_{\mathrm{ini},2}$ e-foldings before the mode exits the Hubble horizon during inflation. After the mode exits the horizon, at $N_{\mathrm{fin}}$, the numerical integration is stopped. For the modes with $k<k_\mathrm{min}$ that never enter the horizon, the stopping time is the same as for $k_\mathrm{min}$.}
\end{figure}

In order to numerically solve the Mukhanov--Sasaki equation \eqref{mode_evolution} for all the relevant modes, we adopt the same strategy employed in Refs.~\cite{Henriques:1993km,Moorhouse:1994nc,Mendes:1994ai,BouhmadiLopez:2009hv,BouhmadiLopez:2011kw,BouhmadiLopez:2012by,BouhmadiLopez:2012qp} and replace the linear perturbation $v_k$ and its derivative by the variables $X_k=v_k/\hbar$ and $Y_k=-(1/\I k)X'/\hbar$. This allows us to re-write the second order linear differential equation as a set of two first order linear differential equations :
\begin{align}
	X_k' = -{\I k}Y_k
	\,,
	\qquad
	Y_k' = - \I k\left(1 - \frac{1}{k^2}\frac{z''}{z}\right)X_k
	\,.
\end{align}
If we further decompose $X_k$ and $Y_k$ into its real and imaginary components $X^\mathrm{(re)}_k$, $X^\mathrm{(im)}_k$, $Y^\mathrm{(re)}_k$ and $Y^\mathrm{(im)}_k$, and change the time variable from the conformal time $\eta$ to the number of e-foldings $N:=\log(a/a_*)$, we obtain the following four-dimensional system of first-order linear differential equations
\begin{align}
	\label{numerical_system}
	\frac{\D}{\D N}
	\begin{pmatrix}
	 X^\mathrm{(re)}_k\\
	 X^\mathrm{(im)}_k\\
	 Y^\mathrm{(re)}_k\\
	 Y^\mathrm{(im)}_k
	\end{pmatrix}
	= 
	\frac{k}{a H}
	\begin{pmatrix}
		0 & 0 & 0 & 1 \\
		0 & 0 & -1 & 0 \\
		0 & \left(1-\frac{1}{k^2}\frac{z''}{z}\right) & 0 & 0 \\
		-\left(1-\frac{1}{k^2}\frac{z''}{z}\right) & 0 & \qquad0\qquad & \qquad0\qquad \\
	\end{pmatrix}
	\cdot
	\begin{pmatrix}
	 X^\mathrm{(re)}_k\\
	 X^\mathrm{(im)}_k\\
	 Y^\mathrm{(re)}_k\\
	 Y^\mathrm{(im)}_k
	\end{pmatrix}
	\,,
\end{align}
subjected to the constraint
\begin{align}
	2k\left(
		X^\mathrm{(re)}_kY^\mathrm{(re)}_k 
		+ X^\mathrm{(im)}_kY^\mathrm{(im)}_k
	\right)
	=1
	\,.
\end{align}

To set the initial conditions for the numerical integration we begin by defining the wave number $k_c$ (cf.~Fig.~\ref{fig_initial conditions}) which characterizes the maximum value of $z''/z$ in the bumps that occur before inflation. Once $k_c$ is properly defined, we set the initial conditions for the integration variables $X^\mathrm{(re)}_k$, $X^\mathrm{(im)}_k$, $Y^\mathrm{(re)}_k$ and $Y^\mathrm{(im)}_k$ according to the following rule:
\begin{itemize}
\item For modes with $k<10^2k_\mathrm{c}$, we set the initial conditions for the perturbations deep inside the kinetically dominated period, at $N=N_{\mathrm{ini},1}$ (cf. the mode $k_1$ in Fig.~\ref{fig_initial conditions}). The values of the integration variables are fixed using the solutions \eqref{solutions_anal} for $w=1$ and setting $c_{1k}=0$ and $c_{2k}=1$.
\item For modes with $k>10^2k_\mathrm{c}$, we consider that the modes are not sensitive to the shape of $z''/z$ during the transition and that they are in the ground state at the beginning of inflation (cf. the mode $k_2$ in Fig.~\ref{fig_initial conditions}). We use the solutions \eqref{solutions_anal} for $w=-1+\alpha/3$, $c_{1k}=1$ and $c_{2k}=0$, to specify the initial values of the integration variables some $ N_{\mathrm{ini},2}$ e-folds before the moment of horizon crossing.
\end{itemize}
The convergence of the numerical solutions is ensured by stopping the numerical integration not at horizon crossing but some $ N_\mathrm{fin}$ e-foldings after the mode has exited the Hubble horizon, as shown in Fig.~\ref{fig_initial conditions}. For modes that verify $k<k_\mathrm{min}$, where $k_\mathrm{min}$ is defined in \eqref{kmin_def} as the minimum of the Hubble horizon in the transition from kinetic dominance to slow-roll inflation, we stop the integration at the same moment as for $k_\mathrm{min}$.

\begin{figure}[t]
\centering
\includegraphics[width=.495\textwidth]{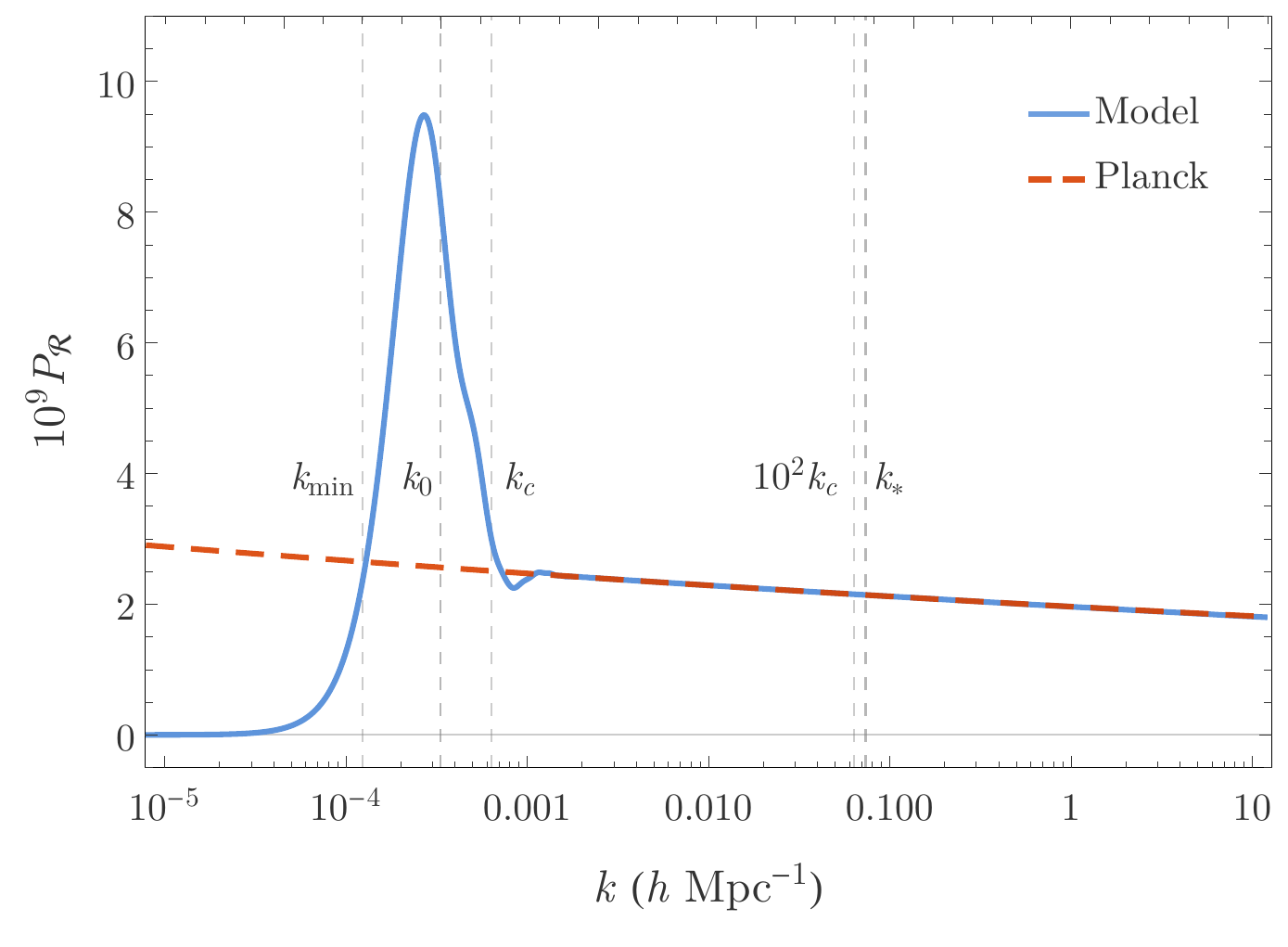}
\hfill
\includegraphics[width=.495\textwidth]{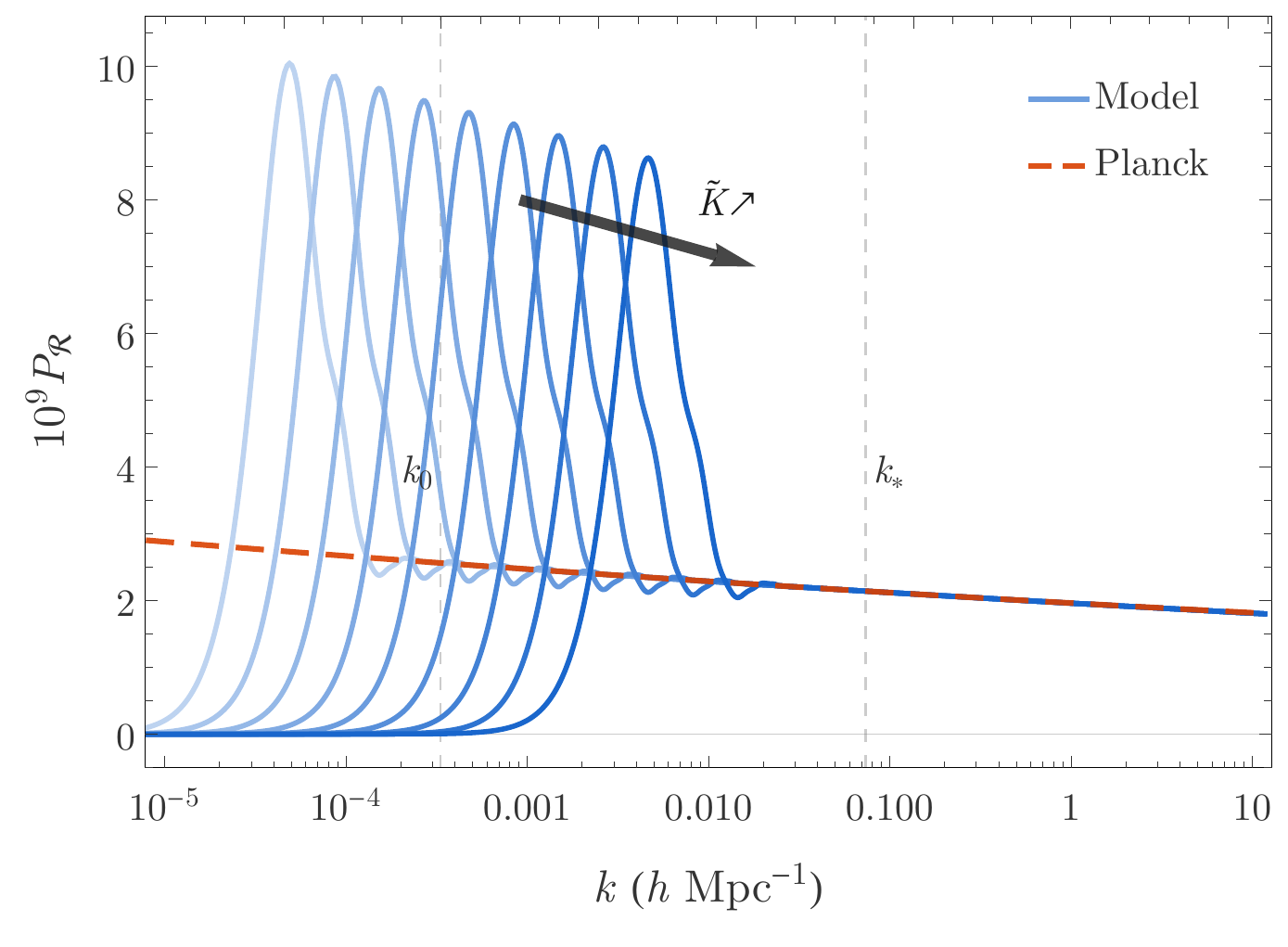}
\caption{\label{fig_spectra} 
(Left Panel) The characteristic power spectrum obtained for the model considered in Eq.~\eqref{bkgd_power_law}. For $k< k_\mathrm{min}$ the power spectrum is highly suppressed while for $k\gg k_\mathrm{c}$ the near scale-invariant shape is recovered. For modes in the range $k_\mathrm{min}\lesssim k \lesssim k_\mathrm{c}$ two peaks of high amplitude appear. The position of these two peaks is related to the amplitude of the potential $z''/z$ in the two bumps visible in Fig.~\ref{fig_initial conditions}. 
For $k\gtrsim 5k_\mathrm{c}$ no imprints of pre-inflationary effects are observed on $P_\mathcal{R}$, validating the upper limit of $k=10^2k_\mathrm{c}$ chosen for setting the initial condition before the onset of inflation.
(Right Panel) The range of modes affected by the pre-inflationary effects is blue shifted as the value of the parameter $\tilde{K}$ increases. In order for any imprints on the power spectrum to be visible while at the same satisfying the observational constraints, the relation \eqref{define_constraint} needs to be satisfied.
}
\end{figure}
On the left-hand side panel of Fig.~\ref{fig_spectra} we present the characteristic shape obtained for the primordial power spectrum\footnote{While in the rest of the paper the wave-number $k$ is dimensionless, in Fig.~\ref{fig_spectra} we follow the convention in the literature and display $k$ in units of $h$ Mpc$^{-1}$.} $P_\mathcal{R}(k)$ (blue) and contrast it with the observational fit (red dashed). 
For modes that verify $k<k_\mathrm{min}$ we observe that $P_\mathcal{R}$ is greatly suppressed. A similar effect can be found in several works that tackle the possibility of finding observable imprints of a pre-inflationary epoch \cite{Contaldi:2003zv,Boyanovsky:2006pm,Piao:2003zm,Piao:2005ag,Liu:2013kea,Powell:2006yg,Wang:2007ws,Scardigli:2010gm,BouhmadiLopez:2012by,Kontou:2017xhp}. In the intermediate range $k_\mathrm{min}\lesssim k \lesssim k_\mathrm{c}$ the power spectrum presents two extremely high peaks that greatly surpass the amplitude of the observational fit. The main peak has an amplitude of $P_\mathcal{R}\sim9\times10^{-9}$ while the secondary one, which appears as a knee near $k_\mathrm{c}$, has an amplitude of $P_\mathcal{R}\sim4.5\times10^{-9}$. Not surprisingly, the position of these peaks corresponds to the wave-numbers that characterize the two bumps of the potential $z''/z$ which can be seen in Fig.~\ref{fig_initial conditions}. For $k\gg k_\mathrm{c}$ no characteristic features appear in the spectrum and the usual near scale-invariant shape is recovered.
The fact that $P_\mathcal{R}$ is smooth in between regions with different initial conditions only validates the choice of the wave number where such a transition occurs. Nevertheless, a lower value than $10^2k_\mathrm{c}$ could be chosen to mark the transition as no imprints due to pre-inflationary effects are found already for $k\gtrsim5k_\mathrm{c}$.
On the right-hand side panel of Fig.~\ref{fig_spectra} we show how the shape of the primordial power spectrum changes for different values of $\tilde{K}$. As $\tilde{K}$ increases (decreases) the range of modes that show imprints of pre-inflationary effects on $P_\mathcal{R}$ is shifted to higher (lower) wave-numbers. Notice that in order for $k_\mathrm{min}>k_0$ to hold, such that the suppression on the large scales affects visible modes, the position of the peaks of the power spectrum starts to approach the pivot scale $k_*$, potentially ruining the viability of the model.

\begin{figure}[t]
\centering
\includegraphics[width=.495\textwidth]{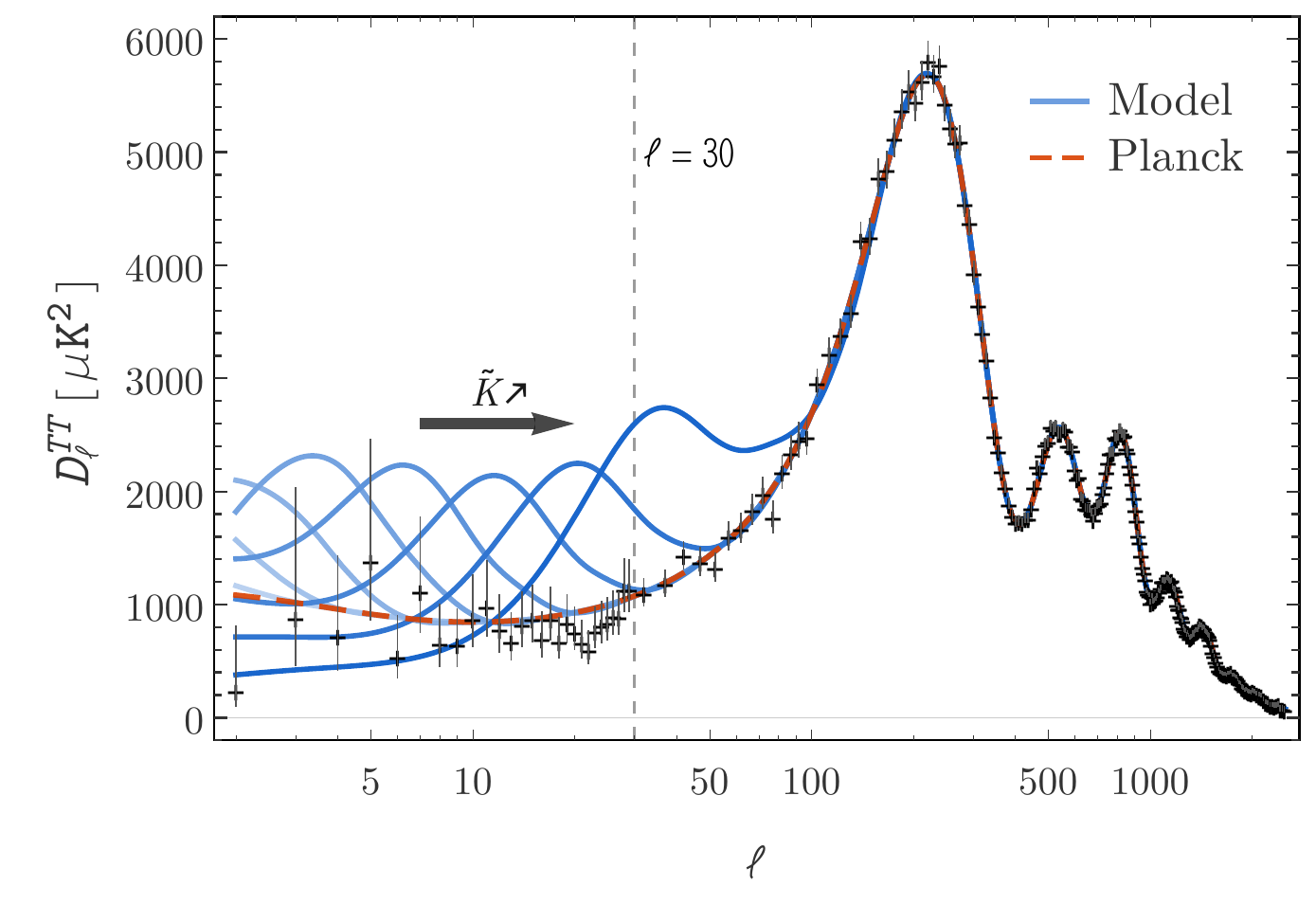}
\hfill
\includegraphics[width=.495\textwidth]{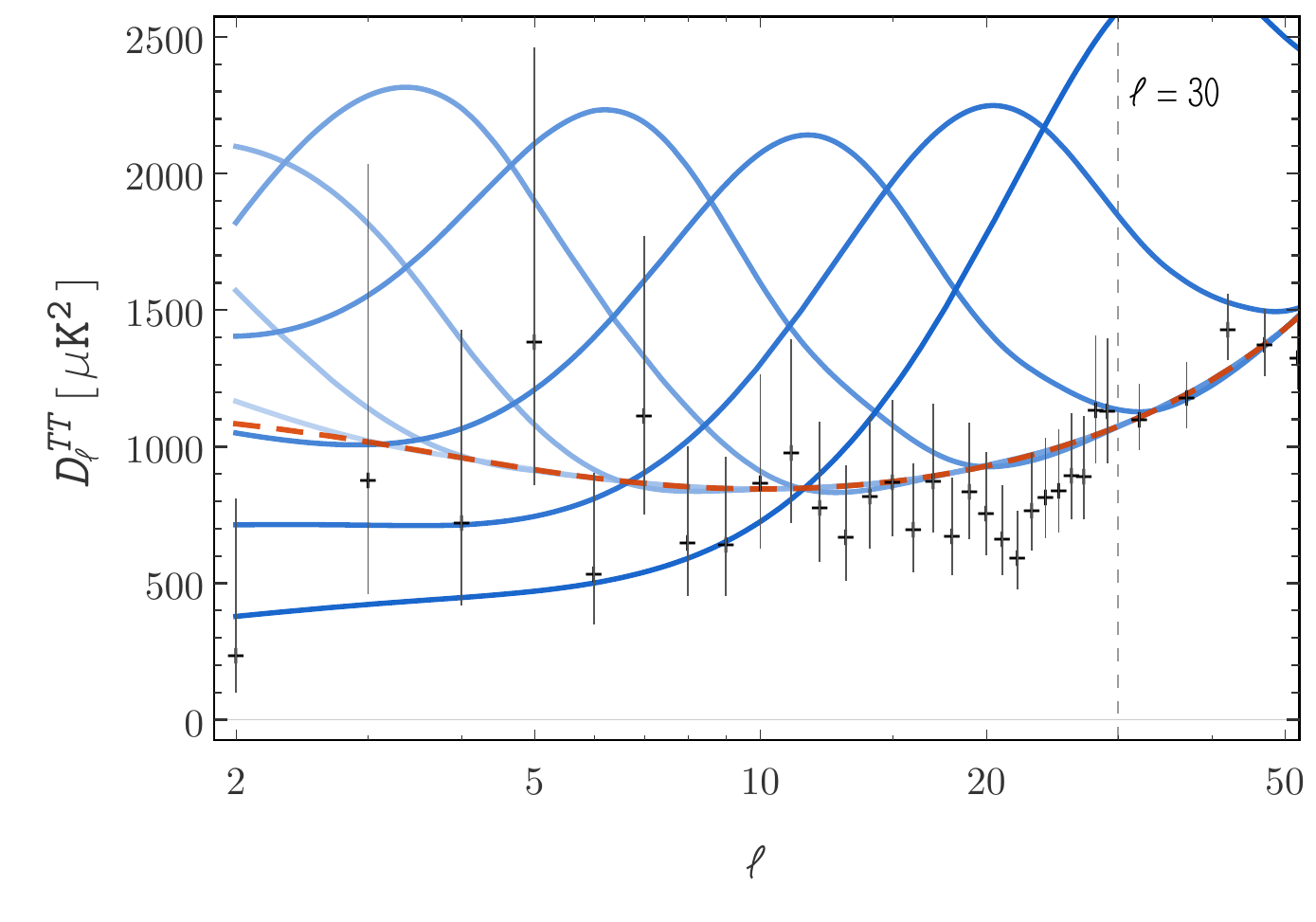}
\caption{\label{fig_Cl} 
The shape of $P_\mathcal{R}$ for the model \eqref{power_spectrum_fit} leads to the suppression of the angular power spectrum $D^{TT}_\ell$ at low multipoles and to the appearance of an extra peak in an intermediate region, while for large multipoles a good fit to the observational data is achieved. The suppression at low multipoles is observed only  for $\tilde{K}\simeq(10k_0/\sqrt{50})^3$ (curve with the third rightmost extra peak). Coincidently, for higher values of $\tilde{K}$ the presence of the extra peak spoils the fit to the observational data for $\ell>30$. The spectra depicted were obtained using the CLASS code \cite{CLASS}.
} 
\end{figure}

In addition to the primordial power spectrum, CMB data constraints on the contribution of the TT-polarization provide a good way to discriminate viable models. The TT contribution to the angular power spectrum from the scalar sector is given by \cite{Ade:2015lrj}
\begin{align}
	C^{TT,s}_\ell:=\int_0^{+\infty}\frac{\D k}{k} \left[\Delta_{\ell,T}^{s}(k)\right]^2P_\mathcal{R}(k)
	\,,
\end{align}
where $\ell$ is the multipole number and the transfer function $\Delta_{\ell,T}^s(k)$ contains all the information necessary from late-time physics.
In Fig.~\ref{fig_Cl} we present the normalized angular power spectra $D^{TT}_\ell:=\ell(\ell+1)C^{TT}_\ell/(2\pi)$ obtained through the use of the CLASS code \cite{CLASS} for (blue) different values of $\tilde{K}$ and (red dashed) the observational fit in \eqref{power_spectrum_fit}. The theoretical predictions are contrasted against the available Planck 2015 data points and error bars \cite{Ade:2015lrj}, to which the red dashed curve presents a remarkable fit for multipoles $\ell\gtrsim30$.
The angular power spectra obtained from the model \eqref{bkgd_power_law} present a strong suppression on the low multipoles as well as a new peak in an intermediate region, while for large multipoles they fit well the observational data. These features can be understood in light of the shape of the the power spectra obtained above -- the suppression on $P_\mathcal{R}$ for $k<k_\mathrm{min}$ leads to the suppression observed in Fig.~\ref{fig_Cl} while the shape of $P_\mathcal{R}$ in the range $k_\mathrm{min}\lesssim k \lesssim \ell_\mathrm{c}$ induces the appearance of the extra peak. As in the case of the primordial power spectrum, the imprints on $D^{TT}_\ell$ are shifted to the right (left) as the value of the parameter $\tilde{K}$ increases (decreases). Despite the possibility of obtaining a suppression on large scales, the existence of the extra peak with such a great amplitude rules out the present model unless the value of $\tilde{K}$ is set so low as to wash away any possible imprints in the visible range. In fact, we note that in order for the suppression on the low multipoles to be observed the good fit to the observational points for $\ell> 30$ has to be spoiled, as can be seen in Fig.~\ref{fig_Cl} from the curve with the third rightmost extra peak, corresponding to $\tilde{K}\simeq(10k_0/\sqrt{50})^3$.

\section{Conclusion}
\label{conclusions}

Despite the incredible fits that inflationary models give to the CMB temperature anisotropies, there are still some anomalies, like the low-quadrupole problem of the CMB, that might hint towards new pre-inflationary physics. In the present paper, we explore the possibility of solving the quadrupole problem in the CMB in the paradigm of the multiverse within the framework of the third quantization. For this goal, we use a toy model as the one presented in the second section and assume, as a first approach, power law inflation despite being aware of the shortcomings of this kind of inflation.

Given our simplified model, it turns out that while we can get a significant suppression of the power spectrum on the largest scales as shown in Fig.~\ref{fig_spectra}, it turns out that generically a new ``bump'' or extra peak appears on the power spectrum between the mode corresponding the pivot scale used in the data analysis of the Planck mission and the scale corresponding to the present Hubble horizon. Therefore, despite the possibility of obtaining a suppression on large scales, the existence of the extra peak with such a great amplitude rules out the present model unless the value of $\tilde{K}$, defined just after Eq.~\eqref{eq214}, is set so low as to wash away any possible imprints in the visible range. Indeed, we point out that in order for the suppression on the low multipoles to be observed, the good fit to the observational points for $\ell> 30$ has to be spoiled, as can be seen in Fig.~\ref{fig_Cl}. A possibility to overcome this problem is to consider an interacting multiverse, for example, as the one presented in \cite{salva2015}, which could alleviate the CMB quadrupole problem along the lines of \cite{Powell:2006yg,Wang:2007ws,Scardigli:2010gm}. We will present those results in a forthcoming paper.

\section*{Acknowledgements}
This article is based upon work from COST Action CA15117 ``Cosmology and Astrophysics Network for Theoretical Advances and Training Actions (CANTATA)'', supported by COST (European Cooperation in Science and Technology). The research of M.\,B.-L.~is supported by the Basque Foundation of Science Ikerbasque. She and J.\,M.~also would like to acknowledge the partial support from the Basque government Grant No.~IT956-16 (Spain) and the project FIS2017-85076-P (MINECO/AEI/FEDER, UE). The research of M.\,K.~was financed by the Polish National Science Center Grant DEC-2012/06/A/ST2/00395 as well as by a grant for the Short Term Scientific Mission (STSM) ``Multiverse impact onto the cosmic microwave background and its relation to modified gravity'' (COST-STSM-CA15117-36137) awarded by the above-mentioned COST Action. For their kind hospitality while part of this work was done, M.\,K.~and J.\,M.~would like to thank the \emph{Centro de Matem\'atica e Aplica\c{c}\~{o}es} of the Universidade da Beira Interior in Covilh\~{a}, Portugal and M.\,K.~also thanks the \emph{Department of Theoretical Physics and History of Science} of the University of the Basque Country (UPV/EHU). M.\,K.~is also grateful to M.\,P.~D\k{a}browski for fruitful discussions. J.\,M.~would like to thank UPV/EHU for a PhD fellowship.

\appendix

\section{Solutions to the Friedmann equation in terms of Hypergeometric Functions}
\label{App_A}

In this section we derive the general solution for the equation
\begin{align}
	\label{app_integral}
	\D T = \frac{\D a}{a^{n}\HdS \sqrt{\dfrac{A}{a^{\alpha}}+\dfrac{B}{a^{6}}}}
	\,,
\end{align}
where $A = a_*^{\alpha}$, $B=\tilde{K}^2/\HdS^6$ , $n>0$ and $\alpha\geq0$. The variable $T$ can represent both the cosmic time ($n=1$) and the conformal time ($n=2$) and the general expression derived from solving Eq.~\eqref{app_integral} includes the solutions of the Friedmann equation in the near de Sitter case of Sec.~\ref{s-model} ($\alpha=0$) and in the power-law inflation of Sec.~\ref{Sec_AsympSols} ($\alpha>0$). 

To integrate the right-hand side of Eq.~\eqref{app_integral} from $0$ to $a_1$ we begin by applying the variable change \mbox{$a\rightarrow z:=(A/B)a^{6-\alpha}$}, after which we obtain for the displacement in $T$
\begin{align}
	\Delta T
	=&~ 
	C
	\int_{0}^{z(a_1)}
	z^{\frac{4-n}{6-\alpha}-1}
	\left(1+z\right)^{\frac{1}{2}-1}
	\D z
	\,,
	\qquad
	C:=
	\frac{1}{\left(6-\alpha\right)\HdS\sqrt{A}}
	\left(\frac{B}{A}\right)^{\frac{1}{2}\frac{2\left(1-n\right)+\alpha}{6-\alpha}}
	\,.
\end{align}
The integral on the right-hand side of the previous equality can be written in terms of an incomplete Beta function \cite{abra,NIST2010} by means of the transformation $z\rightarrow \tilde{z}:=-z$
\begin{align}
	\Delta T
	=&~ 
	-C
	\int_{0}^{\tilde{z}(a_1)}
	(-\tilde{z})^{\frac{4-n}{6-\alpha}-1}
	\left(1-\tilde{z}\right)^{\frac{1}{2}-1}
	\D\tilde{z}
	=
	C\left(-1\right)^{\frac{4-n}{6-\alpha}}
	B_{\tilde{z}(a_1)}\left({\frac{4-n}{6-\alpha}};\,\frac{1}{2}\right)
	\,.
\end{align}
In the last equality we assume the principal value of the factor $(-1)^{\frac{4-n}{6-\alpha}}$. This factor can nevertheless be eliminated by using the hypergeometric representation of the incomplete Beta function (cf. Eq.~(8.17.7) of \cite{NIST2010}),  such that the previous solution becomes
\begin{align}
	\Delta T
	=&~ 
	\frac{a^{4-n}}{\left(4-n\right)\HdS\sqrt{B}}\,
	F\left({\frac{4-n}{6-\alpha}},\,\frac{1}{2};\,{\frac{4-n}{6-\alpha}}+1;\,- \frac{A}{B}a^{6-\alpha}\right)
	\,.
\end{align}
In the particular case of $\alpha=0$  it can be checked that the cosmic time solution \eqref{cosmic_time_sol} is recovered by means of  Eq.~(15.1.7) of \cite{abra} in conjunction with the logarithmic representation of the inverse hyperbolic functions (cf. Eq.~(4.6.20) of \cite{abra}).

\section{Evaluation of the Power Spectrum after horizon crossing}
\label{App2}

In this section we evaluate the primordial power spectrum in power-law inflation outside the Hubble horizon and check whether the formula \eqref{spectrum_walpha} can be extended to the moment of horizon crossing when $k|\eta|\approx1$. 
We begin by defining $a_\mathrm{cross}=a_\mathrm{cross}(k)$ as the value of the scale factor when a certain mode $k$ exits the Hubble horizon and $a_\mathrm{out}:=e^{\Delta N} a_\mathrm{cross}$ as the value of the scale factor when $\Delta N$ e-foldings have passed since that same mode crossed the horizon. During the power-law inflationary regime of \eqref{bkgd_power_law}, the value of the Hubble rate at these two moments is given by
\begin{align}
	H_\mathrm{cross} := H(a_\mathrm{cross}) = \HdS\left(\frac{a_*}{a_\mathrm{cross}}\right)^{\frac{\alpha}{2}}
	\,,
	\qquad
	H_\mathrm{out} := H(a_\mathrm{out}) = \HdS\left(\frac{a_*}{a_\mathrm{out}}\right)^{\frac{\alpha}{2}}
	\,,
\end{align}
so that $a_\mathrm{out}H_\mathrm{out}=k\,\E^{(1-\frac{\alpha}{2})\Delta N}$. Since the argument of the Hankel functions in Eq.~\eqref{solutions_anal} decays exponentially with $\Delta N$, $k|\eta(a_\mathrm{out})|\approx \E^{-(1-\frac{\alpha}{2})\Delta N}$, even a small number of e-foldings after horizon crossing are enough in order for the formula \eqref{spectrum_walpha} to be valid at $a_\mathrm{out}$. This allows us to estimate the primordial power spectrum after horizon crossing as
\begin{align}
	P_{\mathcal{R}}
	\simeq&~
	\frac{1}{\pi\epsilon}\left[{\left(1-\alpha/2\right)}\frac{\Gamma\left(\frac{1}{2}\frac{6-\alpha}{2-\alpha}\right)}{\Gamma(3/2)}\right]^2\frac{\hbar^2 \HdS^2}{\MP^2} \left(\frac{a_*}{a_\mathrm{cross}}\,\E^{-\Delta N}\right)^{\alpha} \left[\frac{\E^{-(1-\frac{\alpha}{2})\Delta N}}{(2-\alpha)}\right]^{-\frac{2\alpha}{2-\alpha}}
	\nn\\
	=&~
	\frac{1}{4\pi\epsilon}\left(\frac{1}{2-\alpha}\right)^{-\frac{2\alpha}{2-\alpha}}\left[{\left(2-\alpha/2\right)}\frac{\Gamma\left(\frac{1}{2}\frac{6-\alpha}{2-\alpha}\right)}{\Gamma(3/2)}\right]^2\frac{\hbar^2 \HdS^2}{\MP^2} \left(\frac{k}{a_*\HdS}\right)^{-\frac{2\alpha}{2-\alpha}}\E^{-\alpha\Delta N} \E^{\alpha\Delta N}
	\nn\\
	=&~
	\frac{1}{2\pi\alpha}\left(\frac{1}{2-\alpha}\right)^{-\frac{2\alpha}{2-\alpha}}\left[{\left(2-\alpha/2\right)}\frac{\Gamma\left(\frac{1}{2}\frac{6-\alpha}{2-\alpha}\right)}{\Gamma(3/2)}\right]^2\frac{\hbar^2 \HdS^2}{\MP^2} \left(\frac{k}{a_*\HdS}\right)^{-\frac{2\alpha}{2-\alpha}}
	\,,
\end{align}
where in the last equality we have used the fact that $\epsilon=\alpha/2$ during power-law inflation. This is the same result found in \eqref{power_spectrum_approx} by directly evaluating \eqref{spectrum_walpha} at horizon crossing and assuming $k_*=a_*\HdS$.


\end{document}